\documentclass[structabstract]{aa}  
\usepackage{epsfig}
\usepackage{epstopdf}
\usepackage{graphicx}
\usepackage{natbib}
\usepackage{txfonts}
\begin{document}
   \title{High Contrast Imaging of the Close Environment of HD\,142527}
   \subtitle{VLT/NaCo adaptive optics thermal and angular differential imaging }
   \author{J. Rameau  \inst{1}
          \and
          G. Chauvin \inst{1}
          \and
          A.-M. Lagrange \inst{1}
          \and
          P. Thebault \inst{2}
          \and
          J. Milli \inst{1}
          \and
          J.H. Girard \inst{3}
          \and
          M. Bonnefoy \inst{4}
          }
   \institute{Institut de Plan\'etologie et d'Astrophysique de Grenoble, UJF, CNRS,
              414 rue de la piscine, 38400 Saint Martin d'H\`eres, France\\
              \email{julien.rameau@obs.ujf-grenoble.fr}
         \and
            LESIA-Observatoire de Paris, CNRS, UMPC Univ. Paris 06, Univ. Paris-Diderot, 92195, Meudon, France
            \and
            European Southern Observatory, Alonso de Cord\'ova 3107, Vitacura, Santiago, Chile
            \and
            Max Planck Institut fur Astronomie K\"onigstuhl 17, D-69117 Heidelberg, Germany
            }
 \thanks{Based on observations collected at the European Organization for Astronomical Research in the Southern Hemisphere, Chile, ESO : run 087.C-0299A.}           
   \date{Received 01 June 2012 ; accepted 06 August 2012}

  \abstract
  {It has long been suggested that circumstellar disks surrounding young stars may be the signposts of planets, and still more since the recent discoveries of embedded substellar companions. The
     planet-disk interaction may create, according to models, large structures, gaps, rings
     or spirals, in the disk. In that sense, the Herbig star
     HD\,142527 is particularly compelling as, its massive disk displays
     intriguing asymmetries that suggest the existence of a dynamical
     peturber of unknown nature. }
   {Our goal was to obtain deep thermal images of the close
     circumstellar environment of HD\,142527 to re-image the reported close-in
     structures (cavity, spiral arms) of the disk and to search for
     stellar and substellar companions that could be connected to their
     presence.}
   {We obtained high contrast images with the NaCo adaptive optics
     system at the Very Large Telescope in L$~\!'$-band. We
     applied different analysis strategies to probe the presence of
     extended structures or point-like sources using both classical
     PSF-subtraction and angular differential imaging.}
      {The circumstellar environment of HD\,142527 is revealed at an unprecedented spatial
       resolution down to the sub arcsecond level for the first time at $3.8~\mu m$. Our images reveal
       important radial and azimuthal asymmetries which invalidate an
       elliptical shape for the disk as previously proposed. It rather suggests a bright inhomogeneous spiral arm plus
       various fainter spiral arms. We also confirm an inner cavity down to $30$~AU and two important dips at position angles of $0$ and
       $135$~deg. The detection performance in angular differential imaging
       enables the exploration of the planetary mass
       regime for projected physical separations as close as $40$~AU. The
       use of our detection map together with Monte Carlo simulations
       sets stringent constraints on the presence of planetary mass,
       brown dwarf or stellar companions as a function of the semi-major axis. They severely constrain the presence of massive giant planets with semi-major axis beyond $50$~AU, i.e. probably within the large disk's cavity that radially extends up
       to $145$~AU or even further outside.}
   {}

   \maketitle

%
\section{Introduction}

Since the discovery of the first exoplanet \citep{mayor95}
around the solar-like star 51 Pegasi, more than 760 have been detected
so far using different observing techniques, from radial velocity
measurements, photometric transit, microlensing to direct imaging.
These discoveries reveal a great diversity of planetary systems in
terms of physical properties (atmosphere and interior) and environment
(single or multiple planets around single stars, in a circumstellar
disk or even circumbinary planets). A new paradigm about the formation, structure and
composition of planets is emerging, wider than what we have learned so
far from the Solar System. Moreover, most of the known detected giant planets
orbit within $5$~AU owing to the detection biases of radial velocity
and transit techniques. The statistical properties of this population
of close-in planets, the period-mass distribution, the
planet-metallicity correlation together with the densities of Hot
Jupiter interiors \citep{udry07, mayor11} favor a
formation mechanism by core accretion followed by gas capture
\citep[CA hereafter,][]{pollack96}. 

At wider orbits, the distribution of giant planets is less well
known. However, the discoveries of planetary mass companions to young, nearby stars in direct
imaging (AB~Pic~b, \citealt{chauvin05}; RXJ\,1609~b, \citealt{lafreniere08}), and further more around Fomalhaut \citep{kalas08} and HR8799 \citep{marois08, marois10} challenge the CA
model. This scenario suffers from exceedingly long timescales and low disk surface density at large separations. Alternative scenarii could operate at wide orbits, either
combined with CA such as planet-planet scattering \citep{crida09} and outward planetary migration \citep[see for instance][]
{morda12} or alternative formation processes must be
invoked like gravitational instability \citep[GI hereafter][]{cameron78} or
stellar binary mechanisms. These formation processes should lead to
different physical and orbital distributions of giant planets \citep{boley09, dodson09}. Hence, new discoveries are needed to understand planet
formation at all orbits.

The technique of direct imaging technique also offers the possibility to detect giant planets in circumstellar disks
to directly study the planet-disk interaction and even dynamically
constrain the planets's mass \citep{chiang09, lagrange12a}. Indirect signs of planetary formation such as clumps, gaps, holes
and spiral arms \citep[see a review in][]{papaloizou07} have been
probably already observed within several proto-planetary disks (e.g AB
Aur, \citealt{fukagawa04}; TCha, \citealt{huelamo11};  \citealt{andrews11}SAO~206462, \citealt{muto12}).

In this context, using high contrast imaging, we have observed the close
environment of the star HD\,142527, known to host an
extended proto-planetary disk.  HD\,142527 is a Herbig star \citet{waelkens96}, classified as a F6IIIe \citep{houk78}. The Hipparcos
measurement has been revisited by \citet{leeuwen07} leading to a parallax of 
$\pi=4.29\pm0.98$~mas, i.e. $d=233^{+69}_{-43}$~pc. However,
HD\,142527 has been identified as a member of the star-forming region
Sco OB2-2 \citep{acke04} or Upper Centaurus Lupus \citep{zeeuw99} hence the distance would be
$d=145\pm15$~pc in both cases. We choose to adopt this latter distance since we
consider the identification of the membership more reliable. The age
and the mass of the star are also matter of debate since
\citet[F06 hereafter]{fukagawa06} derived an age of
$2.2^{+3}_{-2}$~Myr and $M=1.9\pm0.3$~M$_\odot$ from its stellar
UV/optical luminosities and isochrone \citep[from][]{palla99}
whereas \citet{verhoeff11} found an age of $\simeq5$~Myrs and
$M=2.2\pm0.3$~M$_\odot$ based on a new estimation of the stellar
luminosity and pre main sequence isochrones \citep[from][]{siess00}. Finally, \citet{pecaut10} discussed the age of F-type members of the Sco-Cen OB association and estimated it to be in the range $[5;20]~$Myr. We therefore considered an age between $2$ and $20~$Myr for further analysis. The Table~\ref{tab:star_prop} summarizes the stellar
properties of HD\,142527. This system, resolved by F06 in NIR and \citet{fujiwara06} in IR, presents a circumstellar disk quite unusual with
unprecedented infrared excess ($F_{IR}/F_\star=0.98$ and
$F_{NIR}/F_\star=0.32$, \citealt{dominik03}) compared to other Herbig
stars. Evidence of grain growth (deduced from thermal and scattered
light observations) together with the high level of crystallinity (MIDI
observations, \citealt{boekel04}, and NIR spectrum, \citealt{honda09}) reveal that the outer disk is characterized by  highly-processed dust. Moreover, F06 notes the presence of a spiral arm, a
prominent cavity up to $\simeq150$~AU, two asymmetric facing arcs
within the disk and measured a stellar offset of $\simeq 20$~AU with
respect to the center of the disk. Models have been computed to fit the
infrared excess, the geometry of the inner disk and the mineralogy of
the dust \citep{verhoeff11}. They found a disk cavity between $30$
and $130$~AU separating an inner disk embedded in a halo and a
massive outer disk extended to $200$~AU. They conclude that on-going giant planet
formation is suggested from the highly evolved status of the
disk surrounding HD\,142527.

To get a more accurate view of the close-in structures (cavity, spiral
arms) of the disk at unprecedented spatial resolution and to search
for stellar and substellar companions that could be connected to their
formation, we carried out VLT/NaCo observations at L$~\!'$ band using
angular differential imaging (ADI hereafter). We present the observing
strategy and the data reduction in Section \ref{sec:obs}, the fine
structures and the geometry of the cavity and of the outer disk in
Section \ref{sec:disk}, finally the detection performance and the
constraints on the presence of giant planets in Section
\ref{sec:planets}.
 
\begin{table}[th]
\caption{\label{tab:star_prop}HD\,142527 properties}
\centering
\begin{tabular}{lc}
\hline\hline\noalign{\smallskip}
Parameters & \\
\hline\hline\noalign{\smallskip}
RA (J2000)  & $+15:56:41.88$ \\
DEC (J200) & $-42:19:23.27$ \\
Proper motion RA (mas/yr) & $-11.2\pm0.9$\\
Proper motion DEC (mas/yr) & $-24.46\pm0.8$\\
Spectral Type & F6IIIe \\
K (mag) & $4.98\pm0.02$ \\
Distance (pc) & $145\pm 15$ \tablefootmark{a} \\
Age (Myr) & $2-20$ \\
Mass ($M_\odot$) & $2.2\pm0.3$ \tablefootmark {a}\\
\noalign{\smallskip}\hline
\end{tabular}
\tablefoot{
\tablefoottext{a}{Based on new estimation of the stellar luminosity in Verhoeff et al. 2011.}
}
\end{table}

%

 \section{Observations and Data reduction}
\label{sec:obs}
\subsection{Observing strategy and observing conditions}

\begin{table*}[th]
\caption{Observing Log of HD\,142527 in L$~\!'$-band with and
  without the neutral density (ND), as well as for the reference star
  HD\,181296.}
\label{tab:log}
\centering
\begin{tabular}{llllllllll}     
\hline\hline\noalign{\smallskip}
Star & Date & UT-start/end & DIT & NDIT & Nexp & $\pi$-start/end\tablefootmark{a} & $\langle$Airmass$\rangle$\tablefootmark{b} & $\langle \bar\omega\rangle$\tablefootmark{b} & $\langle\tau_0\rangle$\tablefootmark{b} \\
        & &  & (s) & & &($^\circ$) &  & (arcsec) & (ms)\\
\hline\hline\noalign{\smallskip}
HD\,142527 (ND) & 07/29/2011 & 23:49/23:52 & 0.2 & 80 & 10 & -15.4/-12.9 & 1.05 & 0.76 & 1.6 \\
HD\,142527 & 07/29/2011& 23:59/01:03 & 0.2 & 100 & 129 & -7.9/36.5 & 1.05 & 0.55 & 2.1 \\
\noalign{\smallskip}\hline\noalign{\smallskip}
HD\,181296 & 07/29/2011 & 03:23/04:27  & 0.2 & 100 & 124 &  -5.5/23.3  & 1.15 & 1.34 & 0.9 \\
\noalign{\smallskip}\hline
\end{tabular}
\tablefoot{
\tablefoottext{a}{$\pi$ refers to the parallactic angle at the start and the end of observations.}
\tablefoottext{b}{The airmass, the seeing $\bar\omega$ and the coherence time $\tau_0 $ are estimated in real time by the AO system and averaged here over the observing sequence.}
}
\end{table*}

HD\,142527 was observed on July $29^{th}$, 2011 under good seeing
conditions in the course of the thermal imaging survey of 56 young and
dusty stars (Rameau et al. 2012, in prep). We used the VLT/NaCo
\citep{lenzen03, rousset03} high contrast adaptive optics system and imager in angular differential imaging mode \citep[ADI,][]{marois06}. The L$~\!'$-band ($\lambda_0 = 3.8~\mu m$, $\Delta \lambda=0.62~\mu m$) was used with the L27 camera (platescale $\simeq
27.1$~mas/pixel). Twilight flats were obtained for calibration. A
first set of short unsaturated exposures with a neutral density filter
(ND-Long) was acquired to serve as calibration for the point-spread
function (PSF) and for the relative photometry and astrometry
calibrations. A second set of saturated exposures, without the neutral
density filter, was obtained to image the environment of HD\,142527
with a high dynamic range. Both sequences were composed of various offset
positions on the detector for background subtraction and bad or hot
pixels removal. A windowing mode ($512\times514$ pixels) was chosen to
allow short exposure times (DIT) and favor the temporal sampling using
the NaCo cube mode. The parallactic angle variations ($\Delta
\pi=44.4~$deg) are reported on Table~\ref{tab:log}, together with the
exposure time, the number of frames per cube (NDIT), the number of
cubes (Nexp) and the observing conditions. The star HD\,181296 (A0, K =
$5.01~$mag), that was used as reference for classical PSF-subtraction,
was observed with the same observing strategy during the same night.

\subsection{Data processing}

All images were reduced using the pipeline developed for ADI
observations at the Institut de Plan\'etologie et d'Astrophysique de
Grenoble \citep[see][]{lagrange10, bonnefoy11, delorme12,
  chauvin12}. All data were homogeneously flat-fielded and sky-subtracted ; bad and hot pixels were also cleaned from the images. The data were then recentered and bad quality images were removed from the cubes. The result was a master cube containing all individual cleaned and recentered frames, along with their parallactic angles.

To probe the inner part of the environment of HD\,142527, we used
different PSF-subtraction strategies either adapted for the detection
of faint extended structures or to search for point-like
features. These four methods are described below:

\begin{enumerate}
\item Spatial filtering combined with simple derotation (refered as
  nADI). An azimuthal average of the radial profile is estimated and
  subtracted from each image of the master cube, followed by a
  derotation and mean-stacking. A further step consists in an
  additional median filtering (fnADI hereafter).
\item Classical ADI (cADI). The median of all individual reduced images
  is subtracted from the master cube, followed by a derotation and
  mean-stacking of individual images.
\item The LOCI algorithm \citep{lafreniere07}. The residuals are
  minimized for each image using linear combinations of all data to
  derive the PSF contribution at a given location in the image.  For
  LOCI, we considered optimization regions of $N_A = 300~$PSF cores,
  the radial-to-azimuthal width ratio $g=1$, the radial width $\Delta
  r=1$ and $2\times FWHM$, and a separation criteria of $1.0$ and $1.5\times
  FWHM$. A slightly revisited version of LOCI was applied with an
  estimation of the reference that does not take into account the
  contribution of the region of interest (referred as to mLOCI, see \citealt{lagrange12b,milli12}.
\item Finally, classical PSF-subtraction \citep[e.g.][]{mawet09}. The observations were not designed to perform such reduction afterwards but we found that HD\,181296 was observed in our survey (Rameau et al. 2012, in prep.)
  2~hours later with the same parallactic angle conditions. We therefore used it as a reference (see
  table \ref{tab:log} for the observing conditions). To optimally subtract the PSF to HD142527's
    mastercube, we first selected science and reference frames with
    similar parallactic angle variations (covering a total of
    26~deg). From each individual science frame, the closest reference
    one in terms of parallactic angle (within 0.3 deg) was subtracted,
    after recentering and flux rescaling. This step enables the
    optimization the subtraction of the telescope+instrument
    static-speckles. The PSF-subtracted science frames are then
    derotated and mean-stacked. Different methods were applied to
    estimate the flux scaling factor between HD 142527
    and HD 181296 : 1) by applying the same scaling to all frames,
     2) by determining an
    individual scaling factor for each individual frame considering the
    total flux between the science and reference frames, 3/ using two
   independent part of the image(below 0.7 and above 1.5 of radii, to exclude
    the brightest part of the disk) to derive an individual
    scaling factor for each individual frame. The best result was validated according to visual check.
  
  \end{enumerate}

These four PSF-subtraction methods allowed us to reach optimal
detection performance at all separations, identify biases and
non-physical structures. Although ADI (as with high-pass filtering)
processes are well adapted for point-like detections, they are not for
axisymmetric disks with low inclination (in the present case
$i\simeq30^\circ$) since the disks are heavily self-subtracted (Milli
et al. 2012 submitted). However they still enable the detection of
non-axisymmetric structures and sharp edges in the disk. On the other
hand, nADI, low pass filtering and PSF-reference subtraction reveal
large scale features, both in terms of separation and size. Careful
checks have been done with the Airy ring patterns while interpreting
the observed structures. The results of these four methods are
discussed below in the context of the disk morphology and of the
presence of close companions in the environment of HD\,142527.

%
\section{The circumstellar disk around HD\,142527}
\label{sec:disk}

\subsection{Disk morphology}

\begin{figure*}[th]
\centering
\includegraphics[width=9cm]{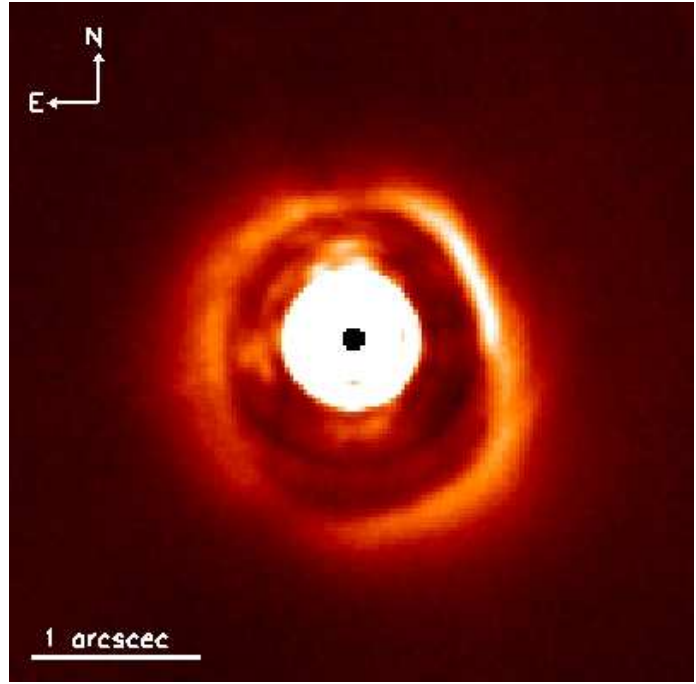}
\caption{HD142527 L$~\!'$ image with a $4.2~\!" \times 4.2 ~\!''$ field of view reduced with the PSF-reference technique and a scaling factor of $2.2$. This factor seems to subtract better the close-in flux without an over subtraction of the most external parts. North is up, East is
  left ; flux scale is linear. }
\label{fig:disk}
\end{figure*}

\begin{figure*}[th]
\centering
\includegraphics[width=16cm]{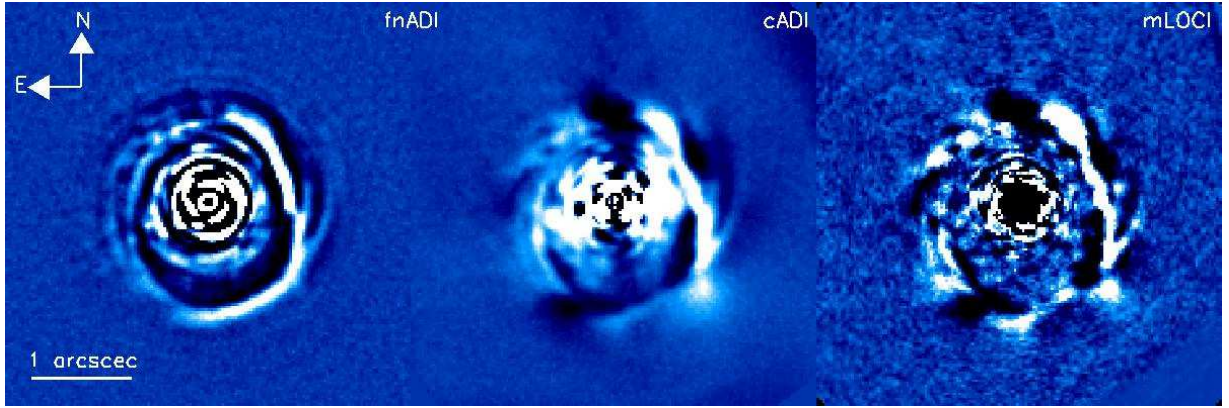}
\caption{HD142527 L$~\!'$ images with a $4.2~\!" \times 4.2 ~\!''$ field of view. North is up, East is
  left. \textbf{Left :} fnADI using a high-pass filtering. \textbf{Middle :} cADI. \textbf{Right :} mLOCI using a
  mask
  to exclude the disk. Note that the linear flux scale is different
  for each panel to reveal all structures. }
\label{fig:adi_disk}
\end{figure*}

Our results on the disk morphology of HD\,142527 are mainly based on
the classical PSF-reference subtraction method, more adapted to the
conservation of faint axisymmetric structures with low
inclination. The complexity of this method is the determination of the
flux scaling factor to avoid the creation of artifacts. After various tests, we ended up with the fact that there
is not an optimal and unique scaling factor for all radii. We therefore tried to
subtract the PSF-reference with a radial scaling factor at each 1 pixel-wide annuli or at
each few pixels-wide annuli. The results were not convincing because
of the asymmetric and high contrasted structures leading to
over-subtracted regions. Unresolved diffuse material seems to 
highly contribute to the PSF-flux observed at L$~\!'$ in addition to the observed structures further away (see
details below).

Nonetheless, it is evident from all images to probe fine highly contrasted
structures. The Figure~\ref{fig:disk} and
Figure~\ref{fig:adi_disk} show the images of HD\,142527 reduced using
the classical PSF-reference subtraction and fnADI, cADI, mLOCI procedures within a reduced FoV of $4.2~\!''
\times 4.2 ~\!''$. However, the photometry of the features remains
sensitive to the choice of the scaling factor. The images reveal new
complex, non-axisymmetrical structures in addition to those previously
reported (F06; \citealt{fujiwara06, ohashi08}). We discuss the main
features below.

\begin{itemize}

\item \textit{Spiral arms}. One of the main results from F06 reveals
  two elliptical facing arcs towards the East-West directions and a displacement of the star with respect to the center of a fitted
  ellipse on the disk. Our data do not confirm this architecture but rather
  suggest spiral shapes. The fnADI image (Figure \ref{fig:adi_disk}, left) remains the most adapted to see the following features. From PA $=170\pm4$~deg to PA $=264\pm2$~deg,
  the arm goes closer to the central star, from $1.17\pm0.04~\!''$ to $0.95\pm0.02~\!''$, with a higher bending at $226~$deg. A remarkable feature
  appears at PA $=265\pm1$~deg : the spiral arm shifts closer
  to the star of $0.11\pm0.02~\!''$. This represents a strong rupture to the
  axisymmetric structure suggested by F06. When going North, the arm moves away (up to $0.91\pm0.03~\!''$). We estimate the arm width of about $0.43~\!''$ for PA $= 227~$deg after PSF-reference subtraction with $sf=2.2$ (without data binning).\\
  Considering the East part of the disk (see Figure \ref{fig:adi_disk}, right), the separation
  increases from North West ($0.87\pm0.03~\!''$) to South East
  ($1.11\pm0.05~\!''$). This spiral arm is not homogenous, showing peaks
  and dips. fnADI, cADI and LOCI highlight this inhomogenity by revealing blobs. We also estimate the arm width for PA$~=47~$deg of about $0.38~\!''$.\\
  Slighlty further out of the main western spiral arm, we detect at least two spiral arms which are born out
   at
  ($1.1\pm0.1~\!''$, PA $=242\pm3$~deg) and
  ($1,01\pm0.04~\!''$, PA $=261\pm3$~deg) and move away while going to
  North. These spirals show up very well in cADI and mLOCI reduced images, up to the $9\sigma$ confidence level
  within the LOCI ones. However, the intensity is not uniform along these arms and their width are narrow
  compared to the ones in the PSF-reference images. These effects due to ADI processing are detailed in Milli et al. (2012, submit.).\\
  Further out from the star, we confirm the presence of a spiral arm between
  ($\simeq1.71~\!''$, $\simeq243$~deg) and
  ($\simeq1.88~\!''$, $\simeq302$~deg) as observed in K and H-band in
  F06. As mentioned in previous papers at shorter wavelengths, we confirm that this arm is fainter with increasing wavelength.
  \item \textit{The brightness asymmetry between the West and East directions}
  It clearly comes out from all images that the West side of the disk is brighter than the East one. This observation confirms the previous one by F06. At the same separation, the Western arm is brighter than the Easter one up to $\simeq2~$mag/arcsec$^2$ (see for instance the SBD profile in Figure\ref{fig:rad_SBD}), bottom).
  \item \textit{The prominent cavity}. We confirm this cavity already observed
  by F06. It is re-detected with an outer edge at
  $0.87\pm0.06~\!''$, i.e. $\simeq 126$~AU at $145$~pc, along
  $PA=227$~deg. F06 presented the cavity to be elliptical and later
  simulations took into account this geometry. However, it is clear
  that the outer edge of the cavity differs from an ellipse. In particular, it is much more pronounced into the South West and extends further up to $145$~AU when going South, as a
  consequence of the spiral shape described earlier. Considering the
  East part, the edge is located at $0.73\pm0.02~\!''$, $PA=64$~deg,
  i.e. $\simeq 106$~AU.
\item \textit{Two azimuthal dips}. We confirm the presence of two big
  dips, detected at North ($PA\in[351\pm2;9\pm5]$~deg and South-East
  ($PA\in[148\pm6;164\pm2]$~deg). The presence of point-like sources in these two dips is discussed below (Section \ref{sec:detlim}). 
   \end{itemize}

\subsection{Surface brightness distribution}

\begin{figure}[th]
\includegraphics[width=9cm]{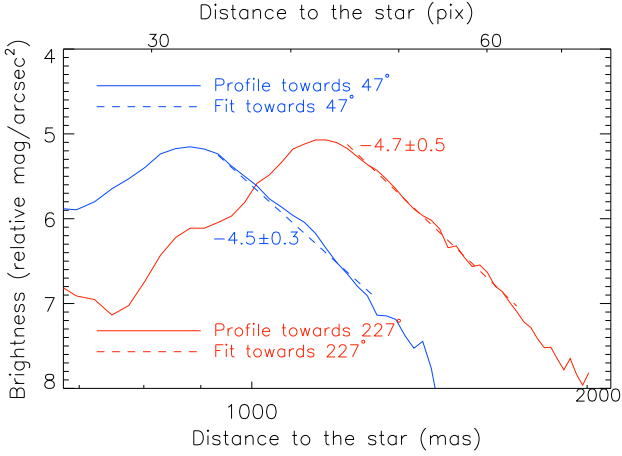}
\centering
\caption{Observed radial surface
  brightness distribution on HD\,142527 L$~\!'$ PSF-reference subtracted image with a scaling factor of $2.2$ towards the
  extended part of the disk ($47-227$~deg). Dashed lines corresponds to fitted straight line
  between $0.9~\!''$ and $1.3~\!''$ for $PA=47$~deg and between
  $1.2~\!''$ and $1.8~\!''$ for $PA=227$~deg.}
\label{fig:rad_SBD}
\end{figure}

We extracted the surface brightness distribution (SBD) of the
HD\,142527's disk using the classical PSF-subtracted image. We
derived the radial profile to evaluate
the extension, the slopes of the outer disk and the amplitude of
features observed in the images towards several PA in $1$ pixel-wide
annulus.  The resulting SBD toward the largest extension of the disk (i.e. PA =$47$ \& $227~$deg) are shown on Figure \ref{fig:rad_SBD}. The slope
 was measured between $0.87~\!''$ and $1.3~\!''$ for PA $=47$~deg and
 between $1.2~\!''$ and $1.8~\!''$ for PA $=227$~deg. We found that
 the SBD follows a simple power low with a slope of
 $\simeq-4.6$. The mass of gas in the disk being poorly constrained
 \citep{ohashi08}, if the circumstellar disk around HD 142527 is
 gas-free disk, such an outer distribution would be indicative of the
 dust dynamics dominated by radiation pressure effects \citep[e.g.][]{thebault08}.
 
To investigate the brightness asymmetry between the West and East side of the disk, we used the GraTeR code \citep{augereau99} to simulate, as a preliminary step, the disk as a ring following a radial volumic distribution of grains : $((r/r_0)^{-2\alpha_{in}}+(r/r_0)^{-2\alpha_{out}})^{-0.5}$. We chose $r_0=145~$AU, $\alpha_{out}=-\alpha_{in}=-3.7$\footnote{The distribution slopes have been chosen to match the SBD slopes measured on the data.}, the disk height scale $\xi=1~$AU at $r_0$, a gaussian vertical profile with $\gamma=2$, a linear disk flaring with $\beta=1$ and an inclination of $i=30~$deg. The disk width thus obtained is $0.48~\!''$, which is larger than the one measured for PA$=227~$deg, but $0.36~\!''$ if we assume $i=45~$deg. We simulated the disk with different scattering properties, from $g=0$ (isotropic) to $g= 0.5$. An anisotropy with $g=0.25$ could explained the flux ratio of $2$ with $i=30~$deg and also $g=0.15$ with $i=45~$deg. However, such configuration could not explain the geometry asymmetry by itself. It might be created either by an offset of the disk center or by an asymmetry between the forward scattering branch and the backward scattering one.

\subsection{Role of a close-in binary companion?}

\citet{biller12} have recently detected a close-in
low-mass star companion candidate, using the NaCo/Sparse Aperture Mode
\citep{tuthill10}. This candidate has a predicted mass of
$0.1-0.4~$M$_\odot$ for a physical projected separation of $88\pm5$mas
($\simeq13~$AU). This candidate should rely at $\simeq3.3~$pixels with a contrast of $5.2~$mag with the primary in our NaCo ADI/L$~\!'$-band
images. However, we did not detect any point source or insight of this companion due to the very small inner working angle and thus numerous residual speckles. If confirmed with a second epoch observation, the companion could be responsible for the
non-axisymmetric features of the circumstellar disk around
HD\,142527. Indeed, the secular perturbations from a central binary
are known to create transient, outward propagating spiral structures
in a circumbinary disc \citep[see for instance the numerical
  investigations of][]{arty94,bate00,deva11}. The extent, evolution
and longevity of these spiral structures depend on several parameters,
such as the binary mass ratio and orbital eccentricity but also the
spatial structure and density of the gas component of the disc. The
detailed study of their evolution in the specific case of HD 142527
requires advanced numerical modeling that clearly exceeds the scope of
the present paper and is left to future investigations.

%
\section{Additional companions around HD\,142527}
\label{sec:planets}

\subsection{Detected point-like sources}

\begin{figure}[th]
\centering
\includegraphics[width=7cm]{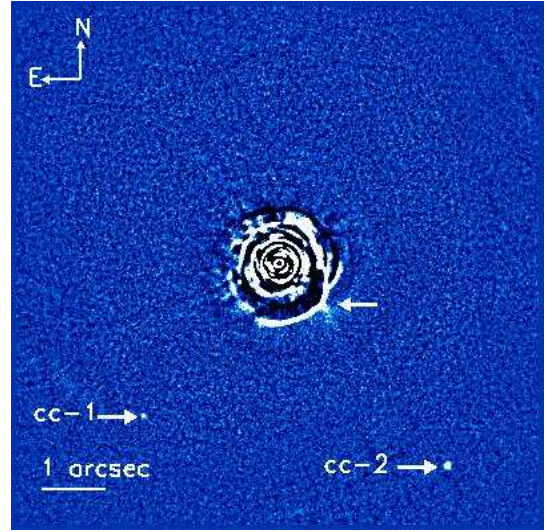}
\caption{fnADI with median filtering image of HD\,142527 with a $11~\!'' \times 11~\!''$ FoV. The arrows indicate candidate companions ; cc-1 and cc-2 are confirmed as background sources whereas the third one is likely an ADI artefact created by the disk. Note that the negative parts (black) close
  to the disk result from the azimuthal averaging.}
\label{fig:bck}
\end{figure}

\begin{table*}[th]
\caption{Relative Astrometry and Photometry of the point-sources detected around HD\,142527.}
\centering
\label{tab:bck}
\renewcommand{\footnoterule}{}
\begin{tabular}{lllllll}   
\hline\hline\noalign{\smallskip}
 Candidate    & UT Date    & Instrument &   Separation       & \textit{PA}  & $\Delta L~\!'$      & Status  \\
              &            &            &   ($''$)            & (deg)        & (mag)              &    \\
\hline\hline\noalign{\smallskip}
cc-1      & 04/06/2004 & Subaru/CIAO& $4.23\pm0.02$     & $140.8\pm0.3$    &               &   \\       
& 07/29/2011 & VLT/NaCo       & $4.168\pm0.008$   & $138.9\pm0.2$    &  $13.3\pm0.3$ &   Background \\ 
cc-2                        & 04/06/2004 & Subaru/CIAO& $5.56\pm0.02$     & $218.5\pm0.3$    &               &  \\  
& 07/29/2011 & VLT/NaCo       & $5.371\pm0.009$   & $219.9\pm0.2$    &  $11.9\pm0.2$ &    Background\\
\hline
\end{tabular}
\end{table*}

We looked for point-like sources in a complete FoV of
$19~\!''\times19~\!''$, obtained after considering all fields observed
during the dithering pattern. We confirm the presence of two faint
point sources (the companion candidates cc-1 and cc-2, see
Figure~\ref{fig:bck}), already detected by F06. Combining the
Subaru/CIAO dataset from 2006 and our NaCo observations, we confirm
the background nature of these two sources. Their relative astrometry
and photometry derived at each epoch are reported in
Table~\ref{tab:bck}.

At closer separations within $2~\!''$, various point-like structures
are detected in the cADI and LOCI images (see
Figure~\ref{fig:disk}). Owing to the low inclination of the disk, the
ADI processing may create artifacts, either extended ones (holes) or
point-like (in the case of abrupt changes in the shape of extended
structures). The bright signal located at $1.3 ~\!''$ and PA
$=227.5$~deg) is for example created by the elbow's shape of the
south-western part of the disk. Consequently, we do not detect any
unambiguous companion candidates within $2~\!''$.

\subsection{Detection limits}
\label{sec:detlim}

\begin{figure}[t]
\centering
\includegraphics[width=10cm]{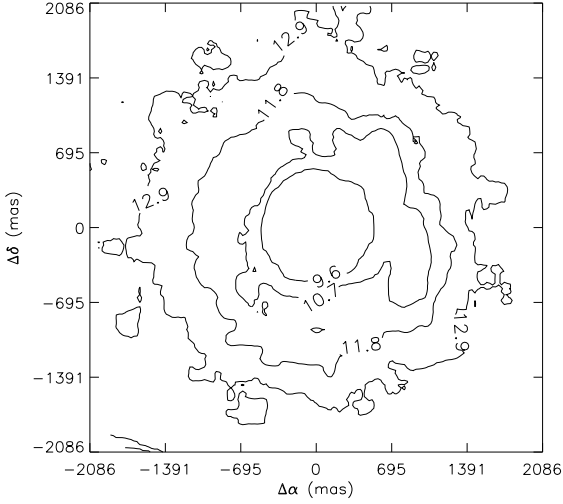}
\caption{2D contrast in L$~\!'$ at $5\sigma$ for
  point-like sources from LOCI reduction around HD\,142527.}
\label{fig:detmap}   
\end{figure}

\begin{figure}[t]
\centering
\includegraphics[width=9cm]{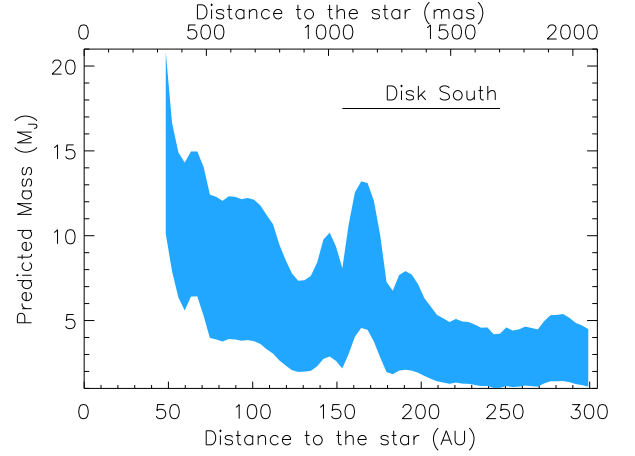}
\caption{Detection limits expressed in Jupiter mass towards the South direction with cADI (similar performance as LOCI). The aqua zone displays the explored masses between $2$ and $20$~Myr since the age of the star is not well established. The position of the disk as been added towards the same PA.}
\label{fig:radlim}   
\end{figure}

To compute the ADI detection performance with LOCI and cADI, the flux
loss of the ADI processing was estimated by injecting fake planets of
given separation and contrast in a cube with empty frames (and in the
data master cube for LOCI). The full pipeline was run. By comparing
the injected fluxes to the recovered ones, we could derive the flux
losses as a function of the separation. The second step was to
estimate the noise in the residual maps using a sliding box of
$9\times9$ pixels. The $5\sigma$ detection limits were then derived by
correcting the noise from the flux loss and applying a normalization
using the unsaturated PSF. We visually checked and confirmed the
detection limits by inserting fake planets with several expected
contrasts at several separations. The 2D-contrast map obtained with
LOCI is shown on Figure \ref{fig:detmap}. We are sensitive to
point-like sources with a contrast of $\Delta L~\!'=10.5$~mag at
$700$~mas ($\Delta_{\rm{proj}} = 101$~AU) and $12.5$~mag at $1500$~mas
(similar performance was obtained with cADI). They correspond to planetary masses of
$6.5~$M$_{\rm{Jup}}$ and $2.5~$M$_{\rm{Jup}}$ at 101 and 218~AU,
respectively, using the predictions of the COND03 evolutionary models
\citep{baraffe03} for an age of $5~$Myrs. We also explored the
performance towards the northern prominent dip of the disk and along
the South direction within the large southern cavity. The 1D predicted mass vs projected separation (only one case is shown because similar
performance is obtained in both directions)
towards the South is reported in Figure \ref{fig:radlim}. Considering
the large uncertainty on the age of the star, whereas the Herbig
status and gas presence are evidence for young age, the explored
masses between $2$ and $20$ Myr are plotted on
the Figure. We do not detect any planets more massive than
$5~$M$_{\rm{Jup}}$ in the northern disk's dip and $4~$M$_{\rm{Jup}}$
in the sourthern cavity. Careful considerations have to be taken since
the detection limits might be overestimated close to the edges of the
disk and to the spiral arms due to the disk impact in ADI processes
\citep[see][]{milli12}.

\subsection{Detection probabilities}

\begin{figure}[th]
\centering
\includegraphics[width=9cm]{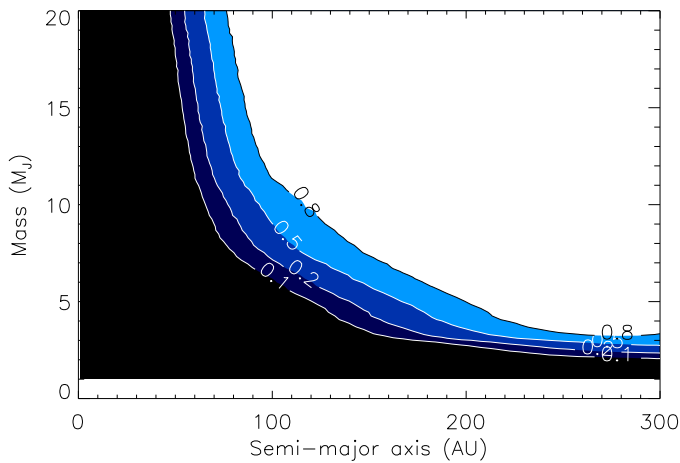} 
\caption{Detection probability map of a faint companion around
  HD\,142527 as a function of mass and semi-major axis. The constraint
  of $i=30$~deg has been set to orbital properties of the generated
  planets. This was calculated using the LOCI 2D detection limits within a $4.2~\!''\times4.2~\!''$ FoV.}
\label{fig:probmap}   
\end{figure}

To set constraints on the physical characteristics of unseen giant
planets or brown dwarfs that could be responsible for several disk
structures, we used the Multi-purpose Exoplanet Simulation System
\citep[the MESS code][]{bonavita12}. Such analysis initiated by \citep{kasper07} has been extensively done up to now, especially to constrain the giant planet population at wide orbits. We first assumed a population of
planets coplanar with the disk by fixing the inclination to 30~deg.
The code creates a planet population using a uniform (mass and
semi-major axis) grid. We explored a parameter space of semi-major
axis ranging between 1 and 500~AU and of masses between 1 and
20~M$_{\rm{Jup}}$, resulting in a final grid of $100\times100$
points. For each point, the code has generated $N_{gen}=10000$ orbits
and computed the projected position of the planet at the tim of the
observation. Using the 2D-contrast maps converted in terms of masses
(using the COND model's predictions, Baraffe et al. 2003) and
projected physical separations, the code finally checked the
detectability of each simulated planet and determined their detection
probability given their masses and semi-major axis (see
Figure~\ref{fig:probmap}. Our NaCo observations set constraints for
planetary mass-objects down to semi-majo axis $a\ge40$~AU and do not
completely explore the large cavity between the outer spiral
structures of HD\,142527 and the inner disk located at less than
30~AU. The detection probabilities unambiguously exclude the presence
of massive planets with M$~\ge9~$M$_{\rm{Jup}}$ and
M$~\ge5$~M$_{\rm{Jup}}$ for semi-major axis of $a\ge130$~AU and
200~AU, respectively. To test the impact of inclination in our
simulations, we ran simulations for $i\in[10,50]$~deg, considering
that a third component will not be necessarly coplanar if the binary
companion detected by Biller et al. (2012) is confirmed. The detection
probabilities do not change significantly our previous results.

\section{Summary and concluding remarks}

Our observations using VLT/NaCo adaptive optics thermal and angular
differential imaging reveal the disk surrounding HD\,142527 at an
unprecedented spatial resolution down to the subarcsecond level. The
disk shows a vast and asymmetric cavity between $30~$AU and $145~$AU
as well as two important dips towards North and South East. We also
resolve a bright and inhomogeneous spiral arm towards the West
direction, several fainter spiral arms and structured features rather
than two elliptical facing arcs as proposed in previous studies. The
observations in angular differential imaging allow us to derive
unprecedented detection performance of point-sources down to
planetary-mass at $40~$AU from the star. Monte-Carlo simulations
combined with our detection map set stringent constraints on the
presence of substellar companions orbiting HD\,142527. We exclude the presence of
brown dwarfs and massive giant planets in the outer part of the cavity
and the presence of giant planets down to $5~$M$_{\rm{Jup}}$ further
out.

The observed structures indicate that the disk is in an active phase
of global evolution. We have already discussed the possibility that
the binary companion candidate might be highly responsible for the
creation of these features. However, other processes can be invoked
such as one or more unseen planets under formation in the
disk.

Gravitational instability is known to create global spiral modes
and in some cases fragmentation. However, given that the
gas mass, as well as the surface density had been poorly constrained so far,
it is not possible to know if the conditions are favorable for gravitational instability. Type II migration of a high mass planet can also be a mechanism able
to produce such structures \citep[e.g.][]{armitage05} in a gaseous
disk. Thus, once again, constraints on the gas properties will
allow to exclude or not the type II migration.

While this article was under review \citet{casassus12} reported Gemini/NICI Ks and L$~\!'$ observations of the outer part of the HD 142527 circumstellar environment. Given similar performance of NICI and NaCo, it gives us the opportunity to validate the observed features and excluded them to be instrumental or analysis artifacts. They confirmed the presence of several structures : the spiral arms, the asymmetric inner cavity, the two North and South South East dips and the inhomogeneity of the East side. However, we do not confirm in our images the presence of what they called a 'knot' along the North West arm. In their images, we also clearly see the offset of the main arm that we reported towards the West, even if they did not mentioned this peculiar morphology. Furthermore, they investigated the presence of a $10~$M$_J$ protoplanet at $90~$AU through hydrodynamic simulations to reproduce the morphology of the inner cavity and of the outer disk. Given our detection performance, such a companion would have likely been detected thus ruling out the input parameters of this carving planet. They argued themselves for additional ingredients within the simulations to reproduce all structures for which one single planet cannot explain. Nevertheless, their approach is very interesting and could be improved by taking into account our detection limits to set planet characteristics and the inner binary candidate companion within FARGO simulations.

From the present data, it is rather difficult to identify the cause of
the non axisymmetric cavity and spiral arms due to the lack of
knowledge on the gas properties. However, we consider that they may be
due to at least one bound-embedded object. Further observations and
important numerical investigations will be necessary to further test
the origin of the highly structured disk of HD\,142527.

 \begin{acknowledgements}

This research has made use of the SIMBAD database and the VizieR
service.  operated at CDS, Strasbourg, France. We would like to thank
M. Bonavita for providing us the MESS code to perform MC simulations. JR also thanks the ESO staff for conducting the observations.
We also acknowledge financial support from the French National
Research Agency (ANR) through project grant ANR10-BLANC0504-01.

\end{acknowledgements}

\nocite*{}
\bibliographystyle{aa}
\bibliography{biblio}

\begin{thebibliography}{61}
\expandafter\ifx\csname natexlab\endcsname\relax\def\natexlab#1{#1}\fi

\bibitem[{{Acke} \& {van den Ancker}(2004)}]{acke04}
{Acke}, B. \& {van den Ancker}, M.~E. 2004, \aap, 426, 151

\bibitem[{{Andrews} {et~al.}(2011){Andrews}, {Wilner}, {Espaillat}, {Hughes},
  {Dullemond}, {McClure}, {Qi}, \& {Brown}}]{andrews11}
{Andrews}, S.~M., {Wilner}, D.~J., {Espaillat}, C., {et~al.} 2011, \apj, 732,
  42

\bibitem[{{Armitage} \& {Rice}(2005)}]{armitage05}
{Armitage}, P.~J. \& {Rice}, W.~K.~M. 2005, ArXiv Astrophysics e-prints

\bibitem[{{Artymowicz} \& {Lubow}(1994)}]{arty94}
{Artymowicz}, P. \& {Lubow}, S.~H. 1994, \apj, 421, 651

\bibitem[{{Augereau} {et~al.}(1999){Augereau}, {Lagrange}, {Mouillet},
  {Papaloizou}, \& {Grorod}}]{augereau99}
{Augereau}, J.~C., {Lagrange}, A.~M., {Mouillet}, D., {Papaloizou}, J.~C.~B.,
  \& {Grorod}, P.~A. 1999, \aap, 348, 557

\bibitem[{{Baraffe} {et~al.}(2003){Baraffe}, {Chabrier}, {Barman}, {Allard}, \&
  {Hauschildt}}]{baraffe03}
{Baraffe}, I., {Chabrier}, G., {Barman}, T.~S., {Allard}, F., \& {Hauschildt},
  P.~H. 2003, \aap, 402, 701

\bibitem[{{Bate}(2000)}]{bate00}
{Bate}, M.~R. 2000, \mnras, 314, 33

\bibitem[{{Biller} {et~al.}(2012){Biller}, {Lacour}, {Juh{\'a}sz}, {Benisty},
  {Chauvin}, {Olofsson}, {Pott}, {M{\"u}ller}, {Sicilia-Aguilar}, {Bonnefoy},
  {Tuthill}, {Thebault}, {Henning}, \& {Crida}}]{biller12}
{Biller}, B., {Lacour}, S., {Juh{\'a}sz}, A., {et~al.} 2012, \apjl, 753, L38

\bibitem[{{Boley}(2009)}]{boley09}
{Boley}, A.~C. 2009, \apjl, 695, L53

\bibitem[{{Bonavita} {et~al.}(2012){Bonavita}, {Chauvin}, {Desidera},
  {Gratton}, {Janson}, {Beuzit}, {Kasper}, \& {Mordasini}}]{bonavita12}
{Bonavita}, M., {Chauvin}, G., {Desidera}, S., {et~al.} 2012, \aap, 537, A67

\bibitem[{{Bonnefoy} {et~al.}(2011){Bonnefoy}, {Lagrange}, {Boccaletti},
  {Chauvin}, {Apai}, {Allard}, {Ehrenreich}, {Girard}, {Mouillet}, {Rouan},
  {Gratadour}, \& {Kasper}}]{bonnefoy11}
{Bonnefoy}, M., {Lagrange}, A.-M., {Boccaletti}, A., {et~al.} 2011, \aap, 528,
  L15+

\bibitem[{{Cameron}(1978)}]{cameron78}
{Cameron}, A.~G.~W. 1978, Moon and Planets, 18, 5

\bibitem[{{Casassus} {et~al.}(2012){Casassus}, {Perez M.}, {Jord{\'a}n},
  {M{\'e}nard}, {Cuadra}, {Schreiber}, {Hales}, \& {Ercolano}}]{casassus12}
{Casassus}, S., {Perez M.}, S., {Jord{\'a}n}, A., {et~al.} 2012, \apjl, 754,
  L31

\bibitem[{{Chauvin} {et~al.}(2012){Chauvin}, {Lagrange}, {Beust}, {Bonnefoy},
  {Boccaletti}, {Apai}, {Allard}, {Ehrenreich}, {Girard}, {Mouillet}, \&
  {Rouan}}]{chauvin12}
{Chauvin}, G., {Lagrange}, A.-M., {Beust}, H., {et~al.} 2012, ArXiv e-prints

\bibitem[{{Chauvin} {et~al.}(2005){Chauvin}, {Lagrange}, {Zuckerman}, {Dumas},
  {Mouillet}, {Song}, {Beuzit}, {Lowrance}, \& {Bessell}}]{chauvin05}
{Chauvin}, G., {Lagrange}, A.-M., {Zuckerman}, B., {et~al.} 2005, \aap, 438,
  L29

\bibitem[{{Chiang} {et~al.}(2009){Chiang}, {Kite}, {Kalas}, {Graham}, \&
  {Clampin}}]{chiang09}
{Chiang}, E., {Kite}, E., {Kalas}, P., {Graham}, J.~R., \& {Clampin}, M. 2009,
  \apj, 693, 734

\bibitem[{{Crida} {et~al.}(2009){Crida}, {Masset}, \& {Morbidelli}}]{crida09}
{Crida}, A., {Masset}, F., \& {Morbidelli}, A. 2009, \apjl, 705, L148

\bibitem[{{de Val-Borro} {et~al.}(2011){de Val-Borro}, {Gahm}, {Stempels}, \&
  {Pepli{\'n}ski}}]{deva11}
{de Val-Borro}, M., {Gahm}, G.~F., {Stempels}, H.~C., \& {Pepli{\'n}ski}, A.
  2011, \mnras, 413, 2679

\bibitem[{{de Zeeuw} {et~al.}(1999){de Zeeuw}, {Hoogerwerf}, {de Bruijne},
  {Brown}, \& {Blaauw}}]{zeeuw99}
{de Zeeuw}, P.~T., {Hoogerwerf}, R., {de Bruijne}, J.~H.~J., {Brown}, A.~G.~A.,
  \& {Blaauw}, A. 1999, \aj, 117, 354

\bibitem[{{Delorme} {et~al.}(2012){Delorme}, {Lagrange}, {Chauvin}, {Bonavita},
  {Lacour}, {Bonnefoy}, {Ehrenreich}, \& {Beust}}]{delorme12}
{Delorme}, P., {Lagrange}, A.~M., {Chauvin}, G., {et~al.} 2012, \aap, 539, A72

\bibitem[{{Dodson-Robinson} {et~al.}(2009){Dodson-Robinson}, {Veras}, {Ford},
  \& {Beichman}}]{dodson09}
{Dodson-Robinson}, S.~E., {Veras}, D., {Ford}, E.~B., \& {Beichman}, C.~A.
  2009, \apj, 707, 79

\bibitem[{{Dominik} {et~al.}(2003){Dominik}, {Dullemond}, {Waters}, \&
  {Walch}}]{dominik03}
{Dominik}, C., {Dullemond}, C.~P., {Waters}, L.~B.~F.~M., \& {Walch}, S. 2003,
  \aap, 398, 607

\bibitem[{{Fujiwara} {et~al.}(2006){Fujiwara}, {Honda}, {Kataza}, {Yamashita},
  {Onaka}, {Fukagawa}, {Okamoto}, {Miyata}, {Sako}, {Fujiyoshi}, \&
  {Sakon}}]{fujiwara06}
{Fujiwara}, H., {Honda}, M., {Kataza}, H., {et~al.} 2006, \apjl, 644, L133

\bibitem[{{Fukagawa} {et~al.}(2004){Fukagawa}, {Hayashi}, {Tamura}, {Itoh},
  {Hayashi}, {Oasa}, {Takeuchi}, {Morino}, {Murakawa}, {Oya}, {Yamashita},
  {Suto}, {Mayama}, {Naoi}, {Ishii}, {Pyo}, {Nishikawa}, {Takato}, {Usuda},
  {Ando}, {Iye}, {Miyama}, \& {Kaifu}}]{fukagawa04}
{Fukagawa}, M., {Hayashi}, M., {Tamura}, M., {et~al.} 2004, \apjl, 605, L53

\bibitem[{{Fukagawa} {et~al.}(2006){Fukagawa}, {Tamura}, {Itoh}, {Kudo},
  {Imaeda}, {Oasa}, {Hayashi}, \& {Hayashi}}]{fukagawa06}
{Fukagawa}, M., {Tamura}, M., {Itoh}, Y., {et~al.} 2006, \apjl, 636, L153

\bibitem[{{Hashimoto} {et~al.}(2011){Hashimoto}, {Tamura}, {Muto}, {Kudo},
  {Fukagawa}, {Fukue}, {Goto}, {Grady}, {Henning}, {Hodapp}, {Honda},
  {Inutsuka}, {Kokubo}, {Knapp}, {McElwain}, {Momose}, {Ohashi}, {Okamoto},
  {Takami}, {Turner}, {Wisniewski}, {Janson}, {Abe}, {Brandner}, {Carson},
  {Egner}, {Feldt}, {Golota}, {Guyon}, {Hayano}, {Hayashi}, {Hayashi}, {Ishii},
  {Kandori}, {Kusakabe}, {Matsuo}, {Mayama}, {Miyama}, {Morino}, {Moro-Martin},
  {Nishimura}, {Pyo}, {Suto}, {Suzuki}, {Takato}, {Terada}, {Thalmann},
  {Tomono}, {Watanabe}, {Yamada}, {Takami}, \& {Usuda}}]{hashimoto11}
{Hashimoto}, J., {Tamura}, M., {Muto}, T., {et~al.} 2011, \apjl, 729, L17

\bibitem[{{Honda} {et~al.}(2009){Honda}, {Inoue}, {Fukagawa}, {Oka},
  {Nakamoto}, {Ishii}, {Terada}, {Takato}, {Kawakita}, {Okamoto}, {Shibai},
  {Tamura}, {Kudo}, \& {Itoh}}]{honda09}
{Honda}, M., {Inoue}, A.~K., {Fukagawa}, M., {et~al.} 2009, \apjl, 690, L110

\bibitem[{{Houk}(1978)}]{houk78}
{Houk}, N. 1978, {Michigan catalogue of two-dimensional spectral types for the
  HD stars}

\bibitem[{{Hu{\'e}lamo} {et~al.}(2011){Hu{\'e}lamo}, {Lacour}, {Tuthill},
  {Ireland}, {Kraus}, \& {Chauvin}}]{huelamo11}
{Hu{\'e}lamo}, N., {Lacour}, S., {Tuthill}, P., {et~al.} 2011, \aap, 528, L7

\bibitem[{{Kalas} {et~al.}(2008){Kalas}, {Graham}, {Chiang}, {Fitzgerald},
  {Clampin}, {Kite}, {Stapelfeldt}, {Marois}, \& {Krist}}]{kalas08}
{Kalas}, P., {Graham}, J.~R., {Chiang}, E., {et~al.} 2008, Science, 322, 1345

\bibitem[{{Kasper} {et~al.}(2007){Kasper}, {Apai}, {Janson}, \&
  {Brandner}}]{kasper07}
{Kasper}, M., {Apai}, D., {Janson}, M., \& {Brandner}, W. 2007, \aap, 472, 321

\bibitem[{{Lafreni{\`e}re} {et~al.}(2008){Lafreni{\`e}re}, {Jayawardhana}, \&
  {van Kerkwijk}}]{lafreniere08}
{Lafreni{\`e}re}, D., {Jayawardhana}, R., \& {van Kerkwijk}, M.~H. 2008, \apjl,
  689, L153

\bibitem[{{Lafreni{\`e}re} {et~al.}(2007){Lafreni{\`e}re}, {Marois}, {Doyon},
  {Nadeau}, \& {Artigau}}]{lafreniere07}
{Lafreni{\`e}re}, D., {Marois}, C., {Doyon}, R., {Nadeau}, D., \& {Artigau},
  {\'E}. 2007, \apj, 660, 770

\bibitem[{{Lagrange} {et~al.}(2010){Lagrange}, {Bonnefoy}, {Chauvin}, {Apai},
  {Ehrenreich}, {Boccaletti}, {Gratadour}, {Rouan}, {Mouillet}, {Lacour}, \&
  {Kasper}}]{lagrange10}
{Lagrange}, A.-M., {Bonnefoy}, M., {Chauvin}, G., {et~al.} 2010, Science, 329,
  57

\bibitem[{{Lagrange} {et~al.}(2012{\natexlab{a}}){Lagrange}, {De Bondt},
  {Meunier}, {Sterzik}, {Beust}, \& {Galland}}]{lagrange12a}
{Lagrange}, A.-M., {De Bondt}, K., {Meunier}, N., {et~al.} 2012{\natexlab{a}},
  ArXiv e-prints

\bibitem[{{Lagrange} {et~al.}(2012{\natexlab{b}}){Lagrange}, {Milli},
  {Boccaletti}, {Lacour}, {Thebault}, {Chauvin}, {Mouillet}, {Augereau},
  {Bonnefoy}, {Ehrenreich}, \& {Kral}}]{lagrange12b}
{Lagrange}, A.-M., {Milli}, J., {Boccaletti}, A., {et~al.} 2012{\natexlab{b}},
  ArXiv e-prints

\bibitem[{{Lenzen} {et~al.}(2003){Lenzen}, {Hartung}, {Brandner}, {Finger},
  {Hubin}, {Lacombe}, {Lagrange}, {Lehnert}, {Moorwood}, \&
  {Mouillet}}]{lenzen03}
{Lenzen}, R., {Hartung}, M., {Brandner}, W., {et~al.} 2003, in SPIE, Vol. 4841,
  944--952

\bibitem[{{Marois} {et~al.}(2006){Marois}, {Lafreni{\`e}re}, {Doyon},
  {Macintosh}, \& {Nadeau}}]{marois06}
{Marois}, C., {Lafreni{\`e}re}, D., {Doyon}, R., {Macintosh}, B., \& {Nadeau},
  D. 2006, \apj, 641, 556

\bibitem[{{Marois} {et~al.}(2008){Marois}, {Macintosh}, {Barman}, {Zuckerman},
  {Song}, {Patience}, {Lafreni{\`e}re}, \& {Doyon}}]{marois08}
{Marois}, C., {Macintosh}, B., {Barman}, T., {et~al.} 2008, Science, 322, 1348

\bibitem[{{Marois} {et~al.}(2010){Marois}, {Zuckerman}, {Konopacky},
  {Macintosh}, \& {Barman}}]{marois10}
{Marois}, C., {Zuckerman}, B., {Konopacky}, Q.~M., {Macintosh}, B., \&
  {Barman}, T. 2010, \nat, 468, 1080

\bibitem[{{Mawet} {et~al.}(2009){Mawet}, {Serabyn}, {Stapelfeldt}, \&
  {Crepp}}]{mawet09}
{Mawet}, D., {Serabyn}, E., {Stapelfeldt}, K., \& {Crepp}, J. 2009, \apjl, 702,
  L47

\bibitem[{{Mayor} {et~al.}(2011){Mayor}, {Marmier}, {Lovis}, {Udry},
  {S{\'e}gransan}, {Pepe}, {Benz}, {Bertaux}, {Bouchy}, {Dumusque}, {Lo Curto},
  {Mordasini}, {Queloz}, \& {Santos}}]{mayor11}
{Mayor}, M., {Marmier}, M., {Lovis}, C., {et~al.} 2011, ArXiv e-prints

\bibitem[{{Mayor} \& {Queloz}(1995)}]{mayor95}
{Mayor}, M. \& {Queloz}, D. 1995, \nat, 378, 355

\bibitem[{{Milli} {et~al.}(2012){Milli}, {Mouillet}, {Lagrange}, {Boccaletti},
  {Mawet}, {Chauvin}, \& {Bonnefoy}}]{milli12}
{Milli}, J., {Mouillet}, D., {Lagrange}, A.~M., {et~al.} 2012, ArXiv e-prints

\bibitem[{{Mordasini} {et~al.}(2012){Mordasini}, {Alibert}, {Benz}, {Klahr}, \&
  {Henning}}]{morda12}
{Mordasini}, C., {Alibert}, Y., {Benz}, W., {Klahr}, H., \& {Henning}, T. 2012,
  \aap, 541, A97

\bibitem[{{Muto} {et~al.}(2012){Muto}, {Grady}, {Hashimoto}, {Fukagawa},
  {Hornbeck}, {Sitko}, {Russell}, {Werren}, {Cur{\'e}}, {Currie}, {Ohashi},
  {Okamoto}, {Momose}, {Honda}, {Inutsuka}, {Takeuchi}, {Dong}, {Abe},
  {Brandner}, {Brandt}, {Carson}, {Egner}, {Feldt}, {Fukue}, {Goto}, {Guyon},
  {Hayano}, {Hayashi}, {Hayashi}, {Henning}, {Hodapp}, {Ishii}, {Iye},
  {Janson}, {Kandori}, {Knapp}, {Kudo}, {Kusakabe}, {Kuzuhara}, {Matsuo},
  {Mayama}, {McElwain}, {Miyama}, {Morino}, {Moro-Martin}, {Nishimura}, {Pyo},
  {Serabyn}, {Suto}, {Suzuki}, {Takami}, {Takato}, {Terada}, {Thalmann},
  {Tomono}, {Turner}, {Watanabe}, {Wisniewski}, {Yamada}, {Takami}, {Usuda}, \&
  {Tamura}}]{muto12}
{Muto}, T., {Grady}, C.~A., {Hashimoto}, J., {et~al.} 2012, \apjl, 748, L22

\bibitem[{{Ohashi}(2008)}]{ohashi08}
{Ohashi}, N. 2008, \apss, 313, 101

\bibitem[{{Palla} \& {Stahler}(1999)}]{palla99}
{Palla}, F. \& {Stahler}, S.~W. 1999, \apj, 525, 772

\bibitem[{{Papaloizou} {et~al.}(2007){Papaloizou}, {Nelson}, {Kley}, {Masset},
  \& {Artymowicz}}]{papaloizou07}
{Papaloizou}, J.~C.~B., {Nelson}, R.~P., {Kley}, W., {Masset}, F.~S., \&
  {Artymowicz}, P. 2007, Protostars and Planets V, 655

\bibitem[{{Pecaut} \& {Mamajek}(2010)}]{pecaut10}
{Pecaut}, M. \& {Mamajek}, E. 2010, in Bulletin of the American Astronomical
  Society, Vol.~42, American Astronomical Society Meeting Abstracts \#215,
   \#455.30

\bibitem[{{Pollack} {et~al.}(1996){Pollack}, {Hubickyj}, {Bodenheimer},
  {Lissauer}, {Podolak}, \& {Greenzweig}}]{pollack96}
{Pollack}, J.~B., {Hubickyj}, O., {Bodenheimer}, P., {et~al.} 1996, \icarus,
  124, 62

\bibitem[{{Rousset} {et~al.}(2003){Rousset}, {Lacombe}, {Puget}, {Hubin},
  {Gendron}, {Fusco}, {Arsenault}, {Charton}, {Feautrier}, {Gigan}, {Kern},
  {Lagrange}, {Madec}, {Mouillet}, {Rabaud}, {Rabou}, {Stadler}, \&
  {Zins}}]{rousset03}
{Rousset}, G., {Lacombe}, F., {Puget}, P., {et~al.} 2003, in SPIE, Vol. 4839,
  140--149

\bibitem[{{Siess} {et~al.}(2000){Siess}, {Dufour}, \& {Forestini}}]{siess00}
{Siess}, L., {Dufour}, E., \& {Forestini}, M. 2000, \aap, 358, 593

\bibitem[{{Th{\'e}bault} \& {Wu}(2008)}]{thebault08}
{Th{\'e}bault}, P. \& {Wu}, Y. 2008, \aap, 481, 713

\bibitem[{{Tuthill} {et~al.}(2010){Tuthill}, {Lacour}, {Amico}, {Ireland},
  {Norris}, {Stewart}, {Evans}, {Kraus}, {Lidman}, {Pompei}, \&
  {Kornweibel}}]{tuthill10}
{Tuthill}, P., {Lacour}, S., {Amico}, P., {et~al.} 2010, in SPIE, Vol. 7735

\bibitem[{{Udry} \& {Santos}(2007)}]{udry07}
{Udry}, S. \& {Santos}, N.~C. 2007, \araa, 45, 397

\bibitem[{{van Boekel} {et~al.}(2004){van Boekel}, {Min}, {Leinert}, {Waters},
  {Richichi}, {Chesneau}, {Dominik}, {Jaffe}, {Dutrey}, {Graser}, {Henning},
  {de Jong}, {K{\"o}hler}, {de Koter}, {Lopez}, {Malbet}, {Morel}, {Paresce},
  {Perrin}, {Preibisch}, {Przygodda}, {Sch{\"o}ller}, \&
  {Wittkowski}}]{boekel04}
{van Boekel}, R., {Min}, M., {Leinert}, C., {et~al.} 2004, \nat, 432, 479

\bibitem[{{van Boekel} {et~al.}(2005){van Boekel}, {Min}, {Waters}, {de Koter},
  {Dominik}, {van den Ancker}, \& {Bouwman}}]{boekel05}
{van Boekel}, R., {Min}, M., {Waters}, L.~B.~F.~M., {et~al.} 2005, \aap, 437,
  189

\bibitem[{{van Leeuwen}(2007)}]{leeuwen07}
{van Leeuwen}, F. 2007, \aap, 474, 653

\bibitem[{{Verhoeff} {et~al.}(2011){Verhoeff}, {Min}, {Pantin}, {Waters},
  {Tielens}, {Honda}, {Fujiwara}, {Bouwman}, {van Boekel}, {Dougherty}, {de
  Koter}, {Dominik}, \& {Mulders}}]{verhoeff11}
{Verhoeff}, A.~P., {Min}, M., {Pantin}, E., {et~al.} 2011, \aap, 528, A91

\bibitem[{{Waelkens} {et~al.}(1996){Waelkens}, {Waters}, {de Graauw}, {Huygen},
  {Malfait}, {Plets}, {Vandenbussche}, {Beintema}, {Boxhoorn}, {Habing},
  {Heras}, {Kester}, {Lahuis}, {Morris}, {Roelfsema}, {Salama}, {Siebenmorgen},
  {Trams}, {van der Bliek}, {Valentijn}, \& {Wesselius}}]{waelkens96}
{Waelkens}, C., {Waters}, L.~B.~F.~M., {de Graauw}, M.~S., {et~al.} 1996, \aap,
  315, L245

\end{thebibliography}

\end{document}